\documentclass[preprint,prl,aps]{revtex4-1}
\usepackage{times,epsf,amsmath,amssymb}
\usepackage{graphics,graphicx,float}
\usepackage[latin1]{inputenc}

\begin{document}
\title{Enhancements to cavity quantum electrodynamics system
\vspace{-6pt}}

\author{A. D. Cimmarusti , J. A. Crawford , D. G. Norris and L. A. Orozco} 
\address{Joint Quantum Institute, Department of Physics, University of Maryland and National Institute of Standards and Technology, College Park, MD 20742-4111, U.S.A.}

\begin{abstract}
We show the planned upgrade of a cavity QED experimental apparatus. The system consists of an optical cavity and an ensemble of ultracold $^{85}$Rb atoms coupled to its mode. We propose enhancements to both. First, we document the building process for a new cavity, with a planned finesse of $\sim$20000. We address problems of maintaining mirror integrity during mounting and improving vibration isolation. Second, we propose improvements to the cold atom source in order to achieve better optical pumping and control over the flux of atoms. We consider a 2-D optical molasses for atomic beam deflection, and show computer simulation results for evaluating the design. We also examine the possibility of all-optical atomic beam focusing, but find that it requires unreasonable experimental parameters.
\end{abstract}

\pacs{42.15.Eq, 42.50.Pq, 32.60.+i \vspace{-4pt}}
\maketitle
\section{Introduction}
The simplest realization of a cavity quantum electrodynamic system consists of a single material fermion coupled to a boson field. Our work in the optical regime couples Rb atoms and a finite number of modes of an optical cavity~\cite{berman94}. These systems have numerous applications in quantum information science~\cite{turchette95b,cirac97,gheri98,wilk07b}, and also enable the study of quantum optics effects difficult to observe in free space~\cite{norris10,norris09a,terraciano09,bishop08,guerlin07,harochebook,hennrich05}.

\begin{figure}[H]
\begin{center}
\includegraphics[width=0.6\textwidth]{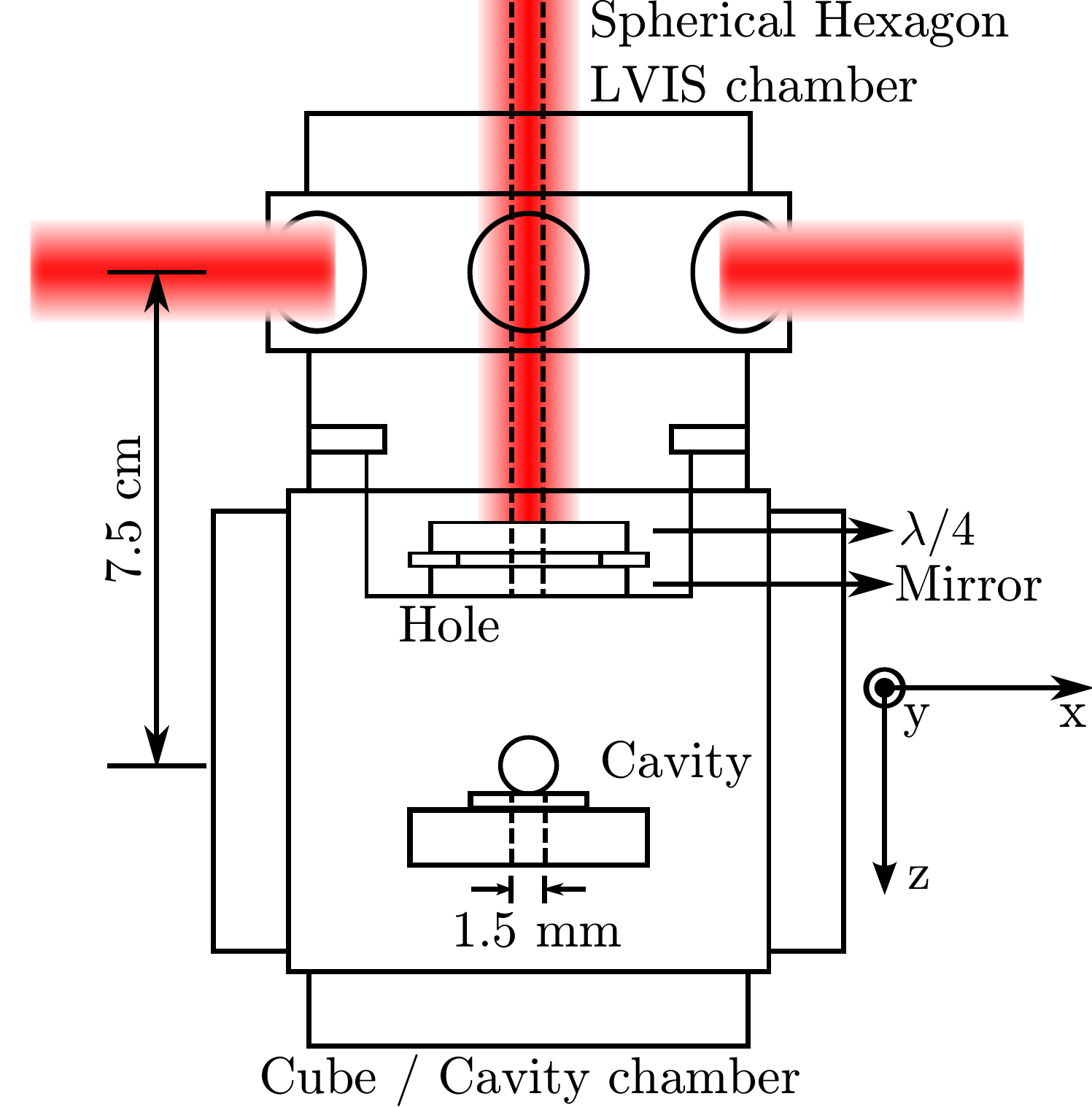}
\caption[chamber]{\label{chamber} (color online) Current vacuum chamber (not drawn to scale). Unbalanced MOT (red beams) by hole in mirror creates LVIS.}
\end{center}
\end{figure}

Our current cavity QED experimental setup incorporates two independent vacuum chamber components: a spherical hexagon and a cube (See Fig.~\ref{chamber}). The former houses an unbalanced Magneto-Optical Trap (MOT), from which we obtain a Low-Velocity Intense Source (LVIS) of $^{85}$Rb atoms~\cite{lu96}. The MOT uses a pair of coils to generate a quadrupole magnetic field, with two retro-reflected laser beams in the horizontal plane, and a third along the axis of the strong magnetic field gradient (vertical direction). At the bottom of the chamber, the vertical beam strikes a gold mirror and quarter-wave plate that have a 1.5-mm diameter hole, yielding a region of unbalanced laser intensity that pushes out a beam of cold atoms. Directly below the hexagon lies a cubic chamber which contains the optical cavity. Atoms propagate 3-4 cm through the mirror to the cavity and are optically pumped to a desired $m$ state \textit{en route}. A fraction of the atoms pass between the cavity mirrors (2.2 mm spacing) and couple to the TEM$_{00}$ mode ($1/e$ field radius 56 $\mu$m) for $\sim$ 5 $\mu$s before striking the bottom of the vacuum chamber.

%figura_(1)

The current arrangement has room for improvement in several fronts: Light from the vertical MOT beam propagates collinearly with the atoms, complicating the desired optical pumping distribution and driving the atoms weakly while they are coupled to the cavity mode. We are unable to access the regime of collective strong coupling in this system, due to the small number of atoms that couple fully to the cavity mode. This is primarily due to the distribution of transverse velocities of the atoms coming from the LVIS, which results in a broadening spatial profile and lower atom density as the beam propagates. To increase the single atom coupling we decrease the mode volume by a factor of two and keep the decay rate of the cavity increasing the finesse also by a factor of two. However, the reduction of the volume complicates the collective strong coupling.

The structure of the paper is as follows: Section 2 talks about the progress in the construction of the new optical cavity. In section 3, we discuss deflection of an LVIS by a 2-D optical molasses (OM). We explore the possibility of focusing an atomic beam in section 4, and we end with conclusions in section 5.
\section{Building the optical cavity}
At the heart of every optical cavity QED apparatus lies a high finesse (low loss) resonator. As the optical transmission coefficients for the mirrors used in the resonator may be as low as a few parts per million, reducing the absorptive losses to this level is a great experimental challenge. Our present cavity (decay rate $\kappa / 2 \pi = 2.8 \times 10^6$ s$^{-1}$  and finesse of 11,000) suffers from higher than expected losses, reducing its finesse by about a factor of two from the intended value ( $\sim$25,000, as calculated using the reflectivity of the mirrors). We believe the losses are due to adhesive residue inadvertently deposited on the mirror surfaces during construction, as well as the aging of the high-reflectivity coatings over the nearly two decades since deposition. We are constructing a new cavity with small and well-characterized losses, as well as slightly stronger coupling to the atoms and enhanced vibration isolation. 

\begin{figure}[H]
\begin{center}
\includegraphics[width=0.85\textwidth]{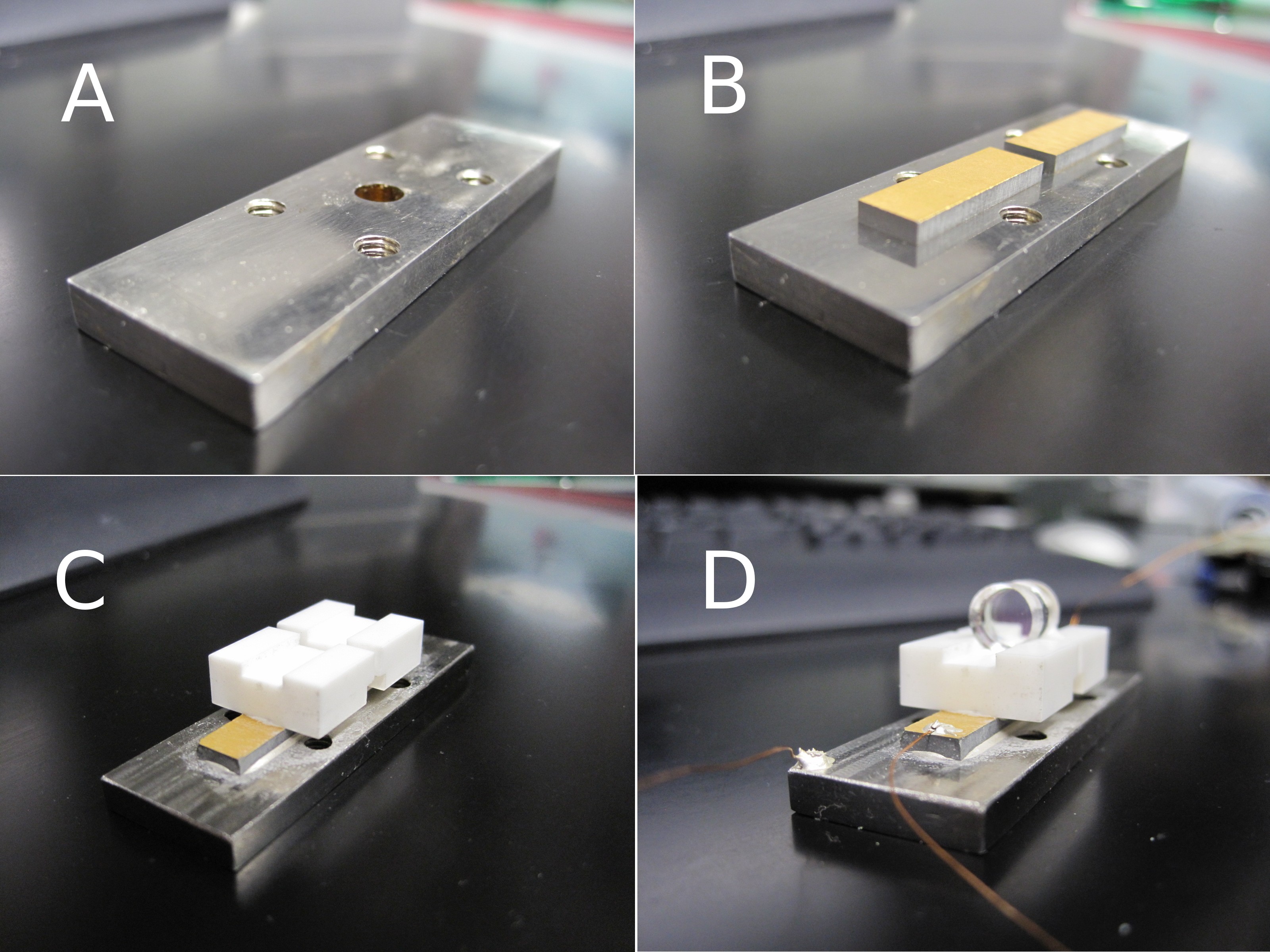}
\caption[Cavity building process]{\label{cavity} (color online) Photographs of the cavity building process. (A) Non-magnetic steel base, (B) PZTs on base, (C) MACOR with slit in the middle and (D) high reflectivity mirrors.}
\end{center}
\end{figure}

Two spherical mirrors with Research Electro-Optics (REO) high-reflectivity coatings form the cavity. For precise alignment, these circular mirrors sit in a square groove machined into a single piece of MACOR, a low-outgassing~\cite{nasa_outgas}, machinable ceramic that can withstand high temperatures. The MACOR is adhered atop two separate piezoelectric transducers (PZTs), bridging the 2 mm gap between them. With the MACOR affixed to the PZTs and the PZTs affixed to a stainless steel base, we cut the MACOR piece in two, thereby allowing for independent translation of each half of the cavity. It is desireable to cut the MACOR with minimal force, so as to avoid putting stress on the PZTs or comprimising the alignment of the groove in the MACOR. This procedure provides good alignment of the MACOR grooves between the two halves, so that simply resting the mirrors in the grooves before gluing keeps the reflective surfaces parallel. To maintain high vacuum ($\sim10^{-8}$ Torr or lower) with the cavity mount inserted in the chamber, we use only adhesives that NASA rates as low-outgassing based on percent total mass loss at low pressure~\cite{nasa_outgas} such as Loctite Hysol 1C. In place of solder, we use EPO-TEK H20E, a silver-filled electrically conductive low-outgassing epoxy, to bond the PZTs to wires and the stainless steel base. Figure~\ref{cavity} shows the various steps followed.

%figura_(2)

The choice of curvature and coating for the two mirrors depends on the desired cavity properties. We aim to construct a cavity with coupling in the intermediate regime (i.e. cavity decay rate $\kappa$ of the same order as single-atom coupling rate $g$ and spontaneous emission rate $\gamma$), and with a finesse of about 20,000. After nearly two decades of aging, the REO mirrors require direct measurement of their transmission properties in order to determine the expected finesse.

We use a 5 mW, 780 nm laser beam and a mounted optical power meter to measure the intensity transmission coefficient $T$ directly. A 780 nm filter reduces background light into the detector to below 0.01 nW. After measuring the power of the unimpeded beam, we move a mirror into the path of the laser such that it is incident on the coated side, and the mirror reflects the majority of the power back through the optical isolator before the laser. Finally, we use an aperture after the mirror to carefully block as much of the scatter as possible (without clipping the transmitted beam itself), and record the intensity of transmitted light. Taking the ratio of the transmitted optical power to the incident optical power (over several independent trials), we determine the current transmissions of the high-reflectivity mirrors. The labeled $T$ values for the mirrors did not always agree with our measurements as Table~\ref{mirrors} shows.

\begin{table}[H]
\begin{center}
\begin{tabular}{|c|c|c|c|c|}
\hline
\# & $R$ & Date & Target $T$ & Range $T$ \\
\hline
6 & 45 cm & 11/98 & 300 ppm & 235 - 297 ppm \\
\hline
3 & 45 cm & 11/98 & 15 ppm & 7.57 - 8.06 ppm \\
\hline
\end{tabular}
\caption[mirror parameters]{\label{mirrors} Mirror parameters: \# number of mirrors, $R$ radius of curvature, Target transmission and Range of trasmission.}
\end{center}
\end{table}

Choosing mirrors with transmissions of $T_1$ = 8 ppm and $T_2 $= 297 ppm (so that over 97\% of photons in the cavity will exit through one mirror, making detection simpler), we calculate finesse assuming that the absorption losses are very small with Eq.~(\ref{finesse}) 
\begin{equation}
\label{finesse}
F = \frac{\pi}{ 1 - \sqrt{ ( 1 - T_1 ) ( 1 - T_2 ) } }.
\end{equation}
These directly measured transmissions give an expected finesse of $\sim$20,000.

We use a method of impedance mismatching in which the stainless steel cavity mount sits atop a stack of materials with very different resonant frequencies, such that vibrations do not easily propagate through the entire stack, to improve the mechanical decoupling of the cavity. The materials include lead, copper, and Sorbothane (a shock absorbing synthetic viscoelastic urethane polymer). In order to find the optimal configuration of the materials, we place the cavity base atop a test stack and use a PZT attached to the base as a microphone, connecting its output to a spectrum analyzer (Stanford Research Systems SR770 FFT Network Analyzer). We then strike the tabletop with a hammer, delivering controlled ``delta function'' impulses, several times over a span of ten seconds. We collect PZT voltage and average the spectrum over a range of frequencies from 0 to 100 kHz, and we compare dozens of different combinations of damping materials.

\begin{figure}[H]
\begin{center}
\includegraphics[width=\textwidth]{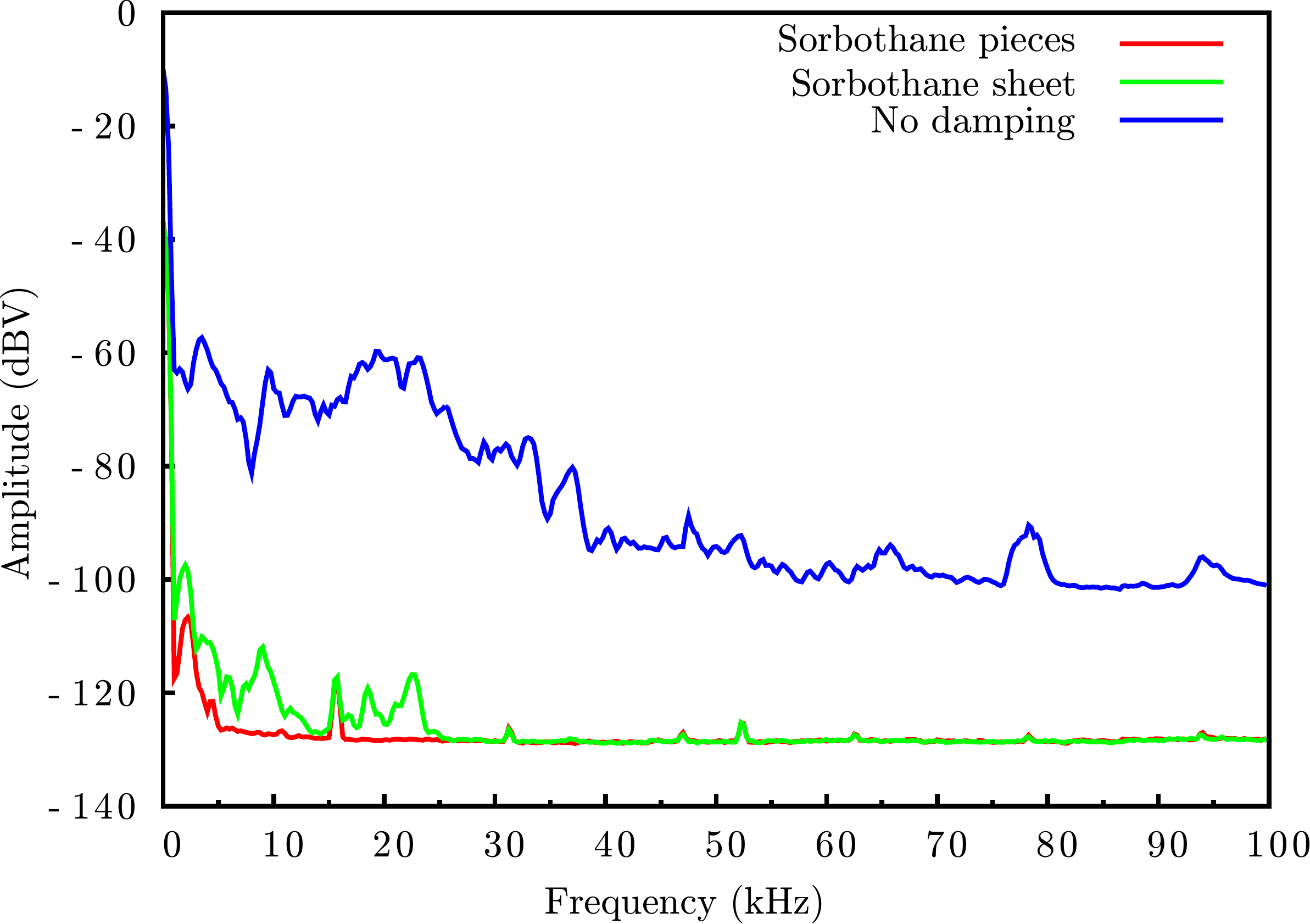}
\caption[vibrations]{\label{vibrations} (color online) Power spectra after impulse excitation. Damping layers: Lead-Sorbothane-Lead.}
\end{center}
\end{figure}

We obtain a drastic improvement to the effectiveness of the damping stack by altering the geometry of the Sorbothane layer, based on advice from the manufacturers of Sorbothane~\cite{sorbothane}. Instead of a solid sheet of Sorbothane, we cut the material into twelve roughly 0.25-inch squares, and space the squares out over the area of the damping stack. As shown in Fig.~\ref{vibrations}, the small pieces of Sorbothane make this geometry significantly more effective at damping, as they have more room to deform sideways and dissipate vibrational energy.

%figura_(3)
%
\section{Atomic beam deflection}
\begin{figure}[H]
\begin{center}
\includegraphics[width=0.6\textwidth]{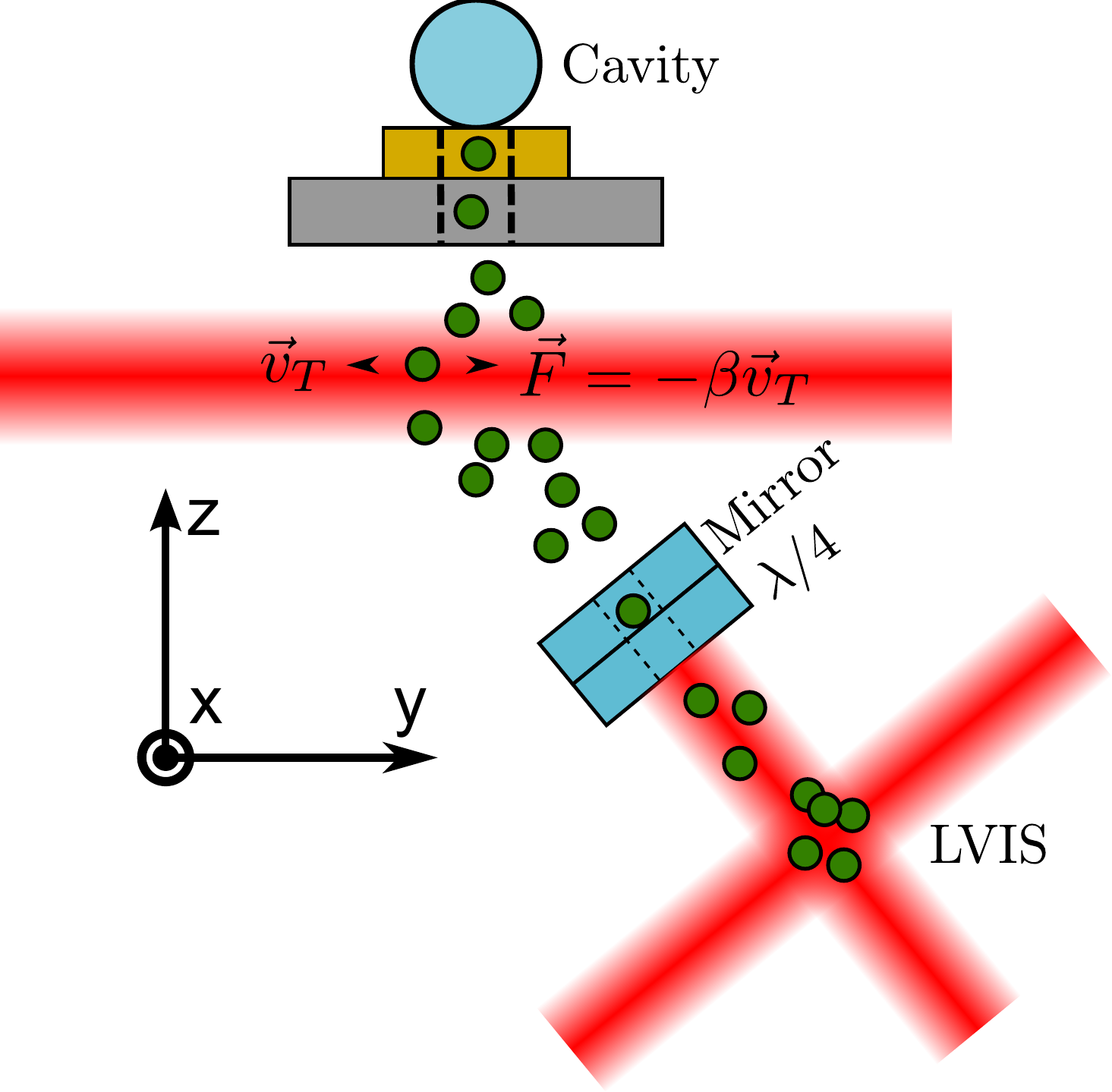}
\caption[molasses deflection]{\label{molasses} (color online) LVIS of atoms exit (a) hole angled to the vertical axis of our optical cavity, scatter light from a (b) 2-D optical molasses.}
\end{center}
\end{figure}

We propose a new atomic beam design to address the problem of the vertical MOT beam that propagates together with the atoms, inspired by ~\cite{wang10,nellessen89,witte92,ashkin70}, in which a 2-D optical molasses deflects the atomic beam from its initial trajectory and into the cavity mode (see Fig.~\ref{molasses}). The benefits of deflection are threefold: It rids us of the unwanted MOT light, it allows for better control of the atom number, and it provides a way to lower the average speed of the atoms traversing the cavity mode. The latter translates into longer transit times for the atoms in the cavity mode.
 
A pair of retro-reflected laser beams, intersecting at right angles in the $x$-$y$ plane, create a 2-D optical molasses (see Fig.~\ref{molasses}). The molasses acts as a viscous damping force of form $\vec{F} = - \beta \vec{v}_T$, opposing the motion of the atoms in the $x$-$y$ plane, and rapidly damping the velocity in these directions, $\vec{v}_T$, to a mean of zero. Aiming the initial atom beam from the LVIS at some angle with respect to the $x$, $y$, and $z$ axes, the molasses damps all velocity components except those along $z$, giving an effective deflection into the $z$ direction while also reducing the mean longitudinal velocity of the beam.

%figura_(4)

Implementing this geometry requires a major modification of our apparatus. We simulate the beam deflection process to find the optimal experimental parameters before changing the system. The code for this simulation is in C++ with the aid of \textsc{Root}, an open source object-oriented data analysis framework developed by CERN and used extensively in particle physics. The code currently relies on histogramming, graphing, geometry and physics libraries of \textsc{Root} and can be obtained from the website of the author under the terms of the \textsc{GNU} General Public License~\cite{andres_sim_beam}
 
The first stage of the simulation involves the modeling of an atomic beam in our current LVIS setup. The main goal of the simulation is to give a reasonable estimate for the ratio of atoms that couple to the cavity mode to the total number of atoms that exit the unbalanced MOT. The code does not simulate a MOT and assumes no atom-atom interactions. 

We create the beam by randomly assigning $(x,y)$ starting coordinates to each atom from a uniform distribution across the area of the hole in the LVIS mirror (see Fig.~\ref{chamber}), defining the vertical starting position as $z = 0$. In a similar way, we randomly specify $(v_x,v_y)$ initial transverse velocities, constrained by the geometric collimation mechanism of the LVIS mirror aperture to a maximum value as in Eq.~(\ref{vtmax}):
\begin{equation}
\label{vtmax}
v_{T,max} = \frac{ D_{h} }{ d_{mh} } v_0,
\end{equation}
where $D_h$ is the diameter of the hole, $d_{mh}$ is the distance from the center of the MOT to the center of the hole and $v_0$ is the mean longitudinal speed of the atom beam from the unbalanced MOT. The initial vertical velocity components $v_z$ come from a Gaussian distribution with mean $v_0 = 14$ m/s and FWHM 2.7 m/s, in accordance with~\cite{lu96}.
 
\begin{figure}[H]
\begin{center}
\includegraphics[width=0.8\textwidth]{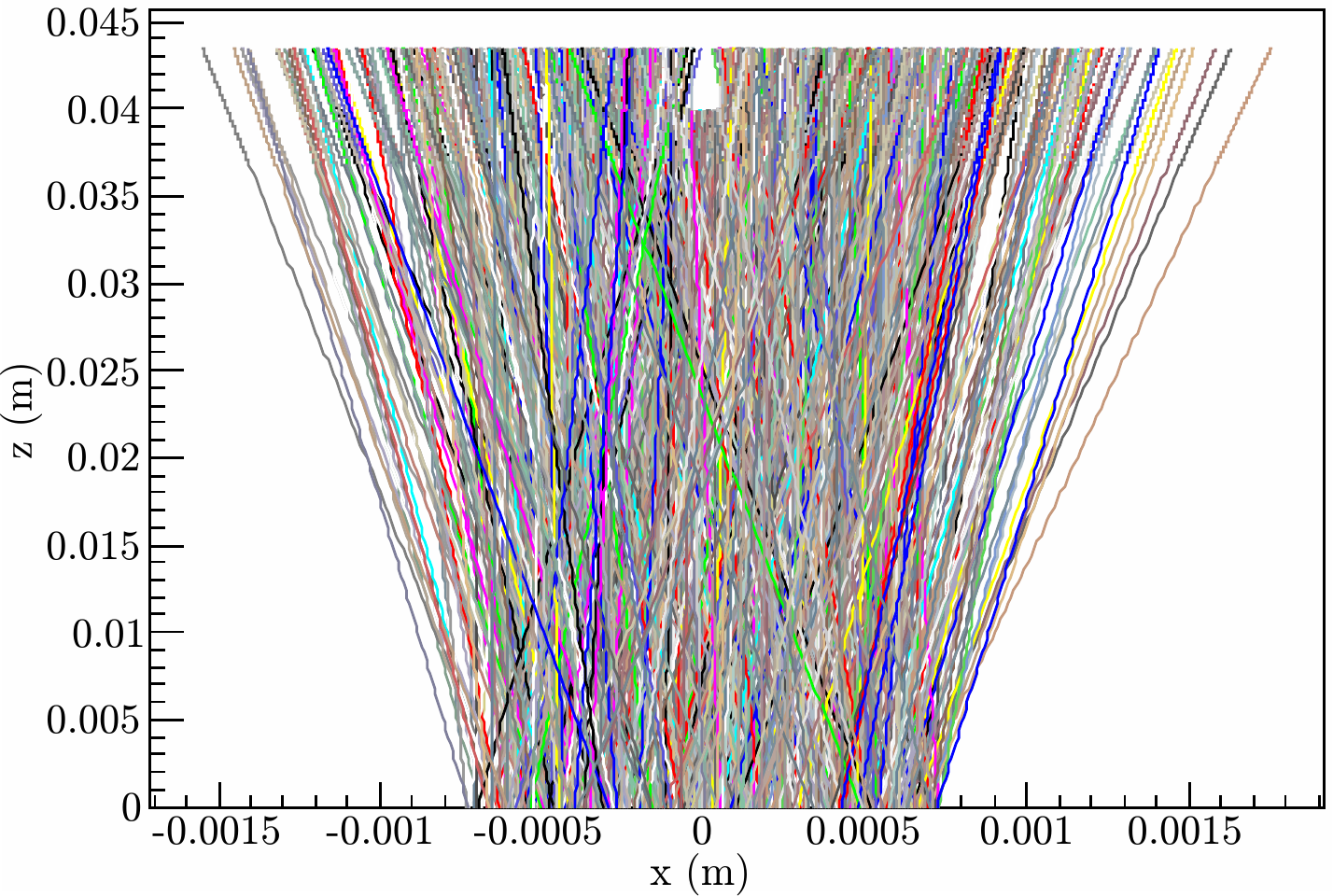}
\caption[vertical atomic beam]{\label{verbeam} (color online) One thousand atomic trajectories a beam exiting a 1.5 mm hole. The optical cavity mirror spacing is 2.2 mm and sits at $z_{cav} = 4$ cm. Colors correspond to different atoms.}
\end{center}
\end{figure}

Figure~\ref{verbeam} shows 1000 atomic trajectories using parameters from our current LVIS and cavity setup. Each atom propagates until it reaches within one waist of the cavity mode center or misses and reaches the chamber wall. The white spot at the top center of the figure corresponds to the position of the cavity mode. The simulations show that 7 - 9 \% of atoms coming from the LVIS couple to the cavity mode, where the range corresponds to the uncertainty in the actual distance between LVIS aperture and cavity mode (between 3 and 4 cm.)  For the planned smaller optical cavity, the range drops to 4 - 6 \%, down by approximately a factor of two as expected from the mode volume.

%figura_(5)

We next simulate the atomic beam deflection proposal. We follow the treatment of~\cite{metcalf99} for the calculation of the scattering rates in steady-state for a two-level atom in the presence of a laser. We give each of the four beams a transverse Gaussian profile as in Eq.~(\ref{gauss}):
\begin{equation}
\label{gauss}
s_{0,i}^{(g)} = s_{0} \left( \frac{w_{0}}{w(z_i)} \right)^2 \exp{ \left[ - 2 \left( \frac{r_i^2}{w^2(z_i)} \right) \right] },
\end{equation}
where $z_i$ and $r_i$ are the longitudinal and transverse coordinates with respect to the $k$-vector of the $i$th beam, $w_0$ is the waist of each beam, $s_{0}$ is the on-resonance saturation parameter at the center of each waist (i.e., at $z_i$=$r_i$= 0), and $w(z_i)$ is the beam spot size as a function of longitudinal position along each beam. $s_{0,i}^{(g)}$ is the Gaussian beam correction to the on-resonance saturation parameter. We use it to calculate the saturation parameter per beam for each atom as Eq.~(\ref{satpar}): 
\begin{equation}
\label{satpar}
s_i = \frac{ s_{0,i}^{(g)} }{ 1 + \left( 2 \left( \delta_i - \vec{k_i} \cdot \vec{v} \right) / \gamma  \right)^2 },
\end{equation}
where $\delta = \omega_{\ell} - \omega_a$ is the laser detuning with respect to the atomic transition frequency, $\vec{k}$ is the laser wavevector, $\vec{v}$ is the velocity of the atom, $\gamma$ is the transition linewidth. We make the approximation that the molasses beams do not interfere, which can be accomplished experimentally by using orthogonal polarizations and spatially separating the $x$ and $y$ beam pairs. In the simplified model we add the contribution from each beam independently and find the total saturation parameter $s_T = \sum_{i = 1}^4 s_i$. We then calculate the total scattering rate $\gamma_p$ as in Eq.~(\ref{scatt}):
\begin{equation}
\label{scatt}
\gamma_p = \frac{ \gamma }{ 2 } \frac{ s_T }{ 1 + s_T },
\end{equation}
When an atom enters the molasses region (defined within the $1/e^2$ intensity region of the beams), it immediately begins scattering at a rate $\gamma_p$. A scattering event involves a recoil velocity of magnitude $\hbar k/m$ in the direction of the absorbed beam, and a recoil of the same magnitude but random direction for the reemission. We determine which beam the atom absorbs from by comparing a random number to the ratio $\gamma_{p,i} / \gamma_p$, \textit{i.e.} the relative weight of each beam compared to the total scattering rate. The beam with the largest weight will be most likely to give the atom a momentum kick. We then select two more random numbers to determine the angle of reemission in three dimensions. Updating the velocity of the atom with the result of the two recoils, we allow the atom to propagate freely until the next scattering event. To determine this time step, we pick random numbers from the exponential waiting-time distribution (Eq.~\ref{wait}),
\begin{equation}
\label{wait}
W(t) = \gamma_p e^{ - \gamma_p t },
\end{equation}
which governs the probability of the next event occuring in time $t$ for a Poisson random process. At each time step we record the position and velocity of the atom. The algorithm repeats until the atom exits the molasses. 

We use very weakly focused Gaussian beams with parameters $w_0 = w(z_i) = 7.5$ mm, for simulating the proposed atomic beam setup, such that the beam intensity has negligible $z_i$ dependence. We choose a slightly smaller aperture size in the LVIS mirror, 1 mm, to achieve a smaller initial beam, and we set the angles of the initial LVIS beam trajectory as $\theta = 30^{\circ}$ (polar) and $\phi = 45^{\circ}$ (azimuthal) with respect to the cavity coordinate system. 

\begin{figure}[H]
\begin{center}
\includegraphics[width=0.8\textwidth]{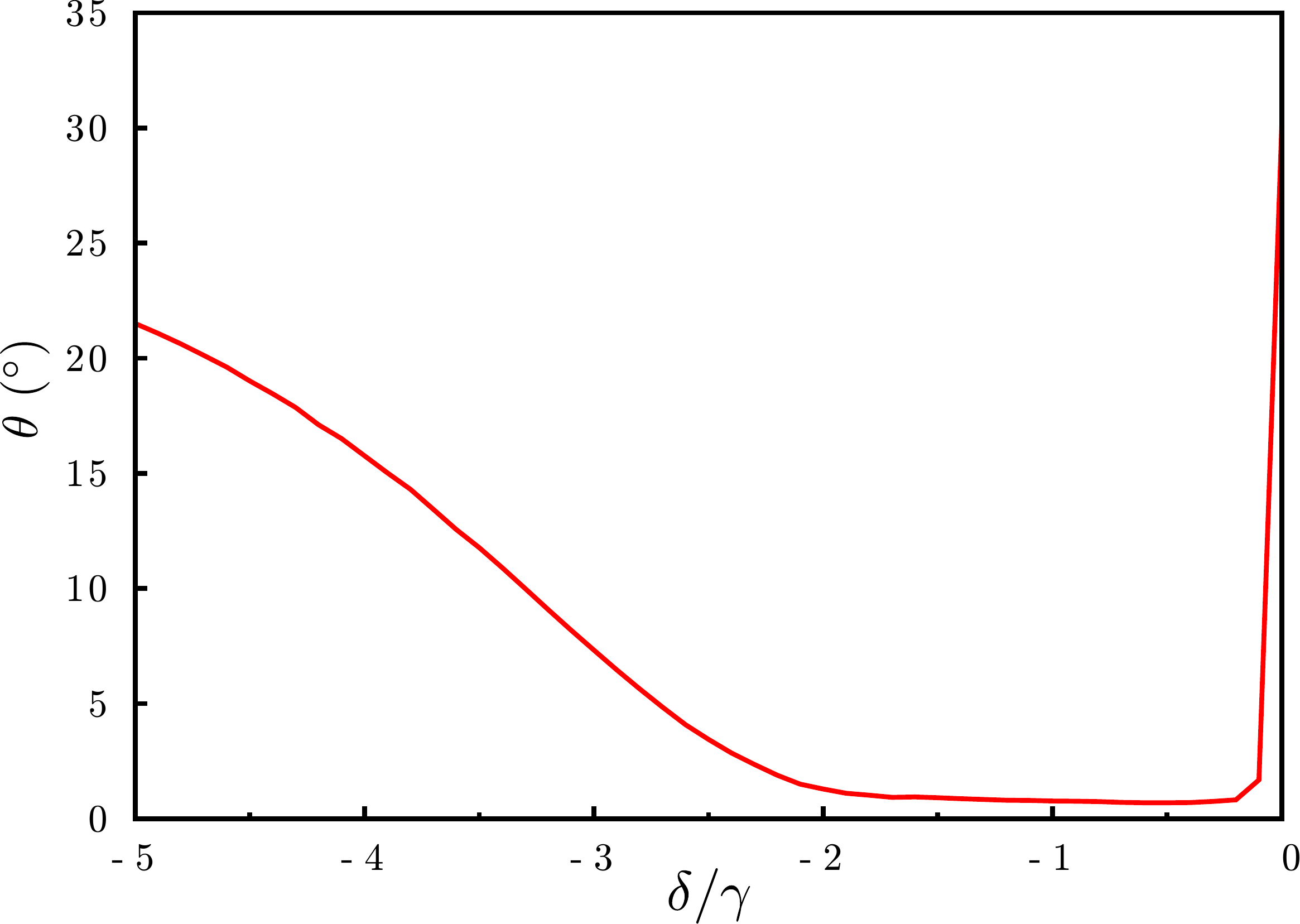}
\caption[detuning exploration]{\label{detun_mol} Average polar angle of atoms after 2-D OM as a function of laser detuning for beams with $w_0 = w = 7.5$ mm and $s_0 = 3$.}
\end{center}
\end{figure}

Figure~\ref{detun_mol} shows the effect of laser detuning on the beam profile. We graph the average polar angle of the atomic beam velocity at a fixed position just after it exits the molasses region. 

%figura_(6)

\begin{figure}[H]
\begin{center}
\includegraphics[width=0.8\textwidth]{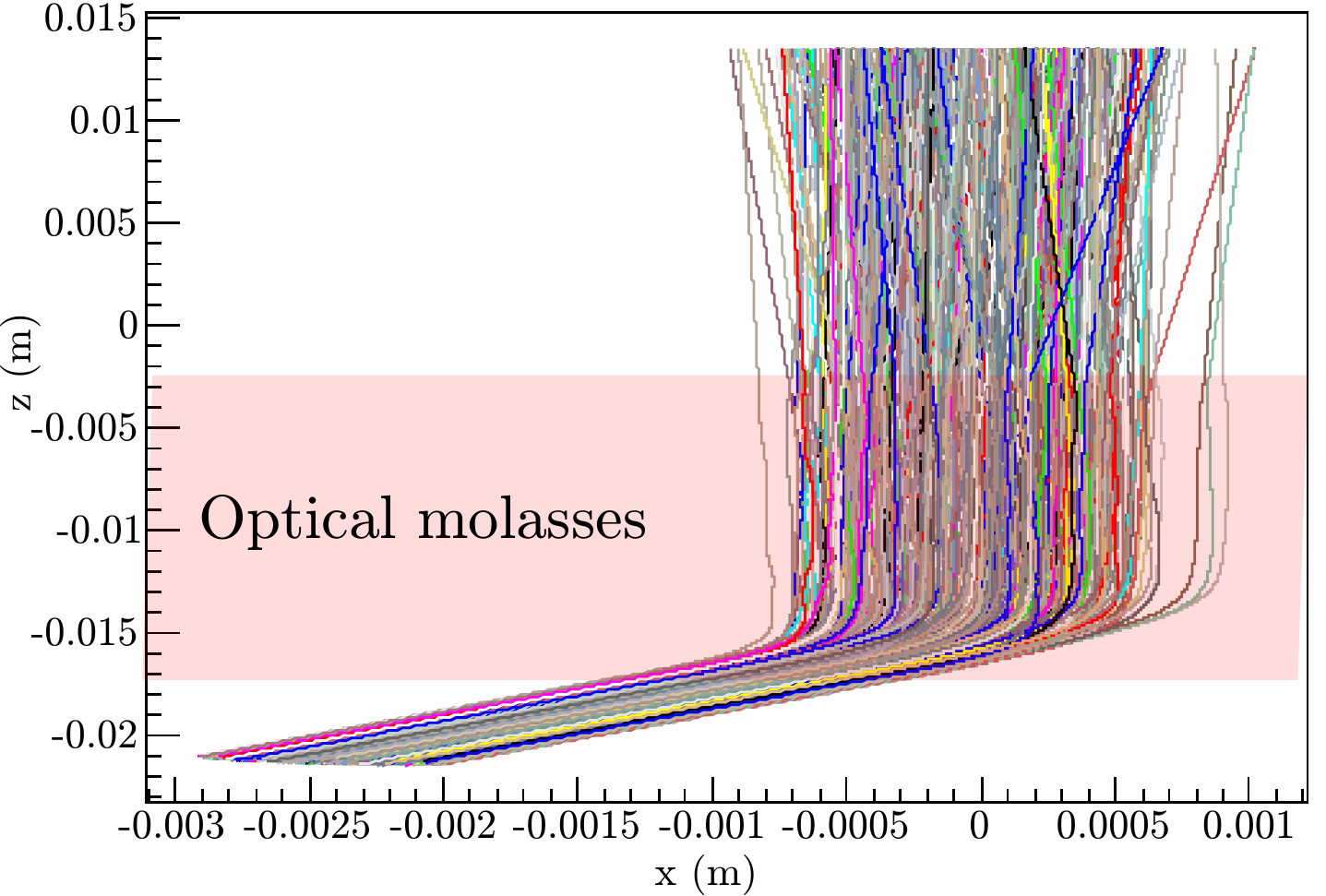}
\caption[atomic trajectories molasses]{\label{zvsx_mol} (color online) Atomic beam trajectories deflected by 2-D OM with $\delta / \gamma = -0.5$ and $s_0 = 3$.}
\end{center}
\end{figure}

Using the same parameters as before, but choosing the optimal detuning of $\delta / \gamma = -0.5$, Fig.~\ref{zvsx_mol} shows atomic trajectories under the action of the 2-D optical molasses. For this simulation, we placed the cavity at $z_{cav} = 1$ cm and obtained 11 \% of atoms coupling to the mode.

%figura_(7)

\begin{figure}[H]
\begin{center}
\includegraphics[width=0.8\textwidth]{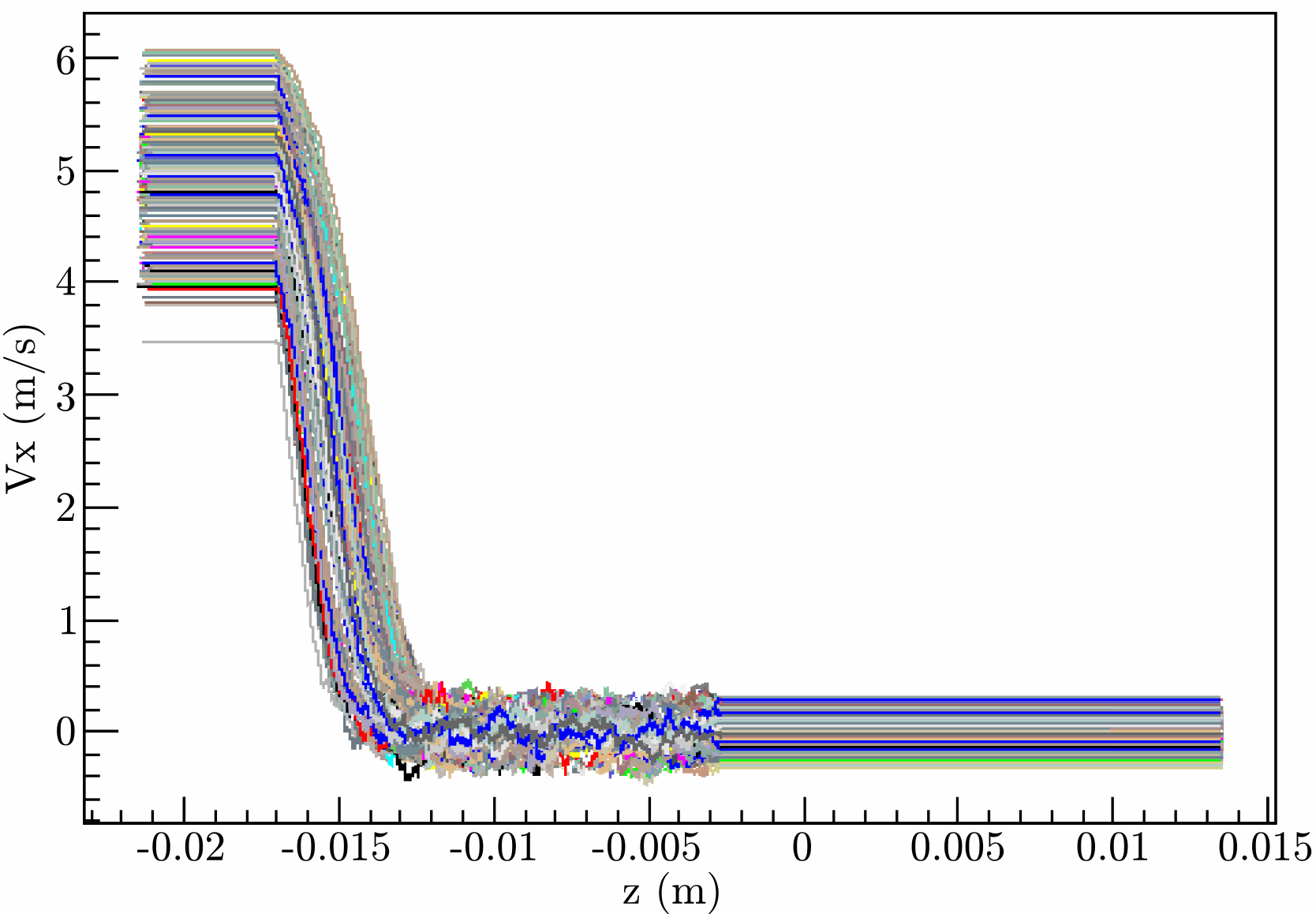}
\caption[atomic velocities molasses]{\label{vxvsz_mol} (color online) Atomic beam $v_x$ damped by 2-D OM with $\delta / \gamma = -0.5$ and $s_0 = 3$.}
\end{center}
\end{figure}

Figure~\ref{vxvsz_mol} shows the $x$ component of the velocity for the atoms. We calculate the RMS transverse speed of the collimated atomic beam and find 15.3 cm/s, yielding a Doppler temperature of 120 $\mu$K. This value agrees with a simple calculation. For a true 1-D system with two beams, we write the heating rate as $4 \gamma_p E_r$, where $E_r = \hbar \omega_r$ is the recoil energy~\cite{metcalf99}. One factor of two comes from the presence of two beams and the second because each absorption and re-emission cycle involves two recoils in that dimension. In three dimensions, the recoil energy from isotropic re-emission is divided equally among all directions, so this second factor of two becomes $( 1 + N/3 )$, where $N$ is the number of dimensions with a pair of molasses beams. When equated to the cooling rate along any one dimension, this gives Doppler temperature $T = \hbar ( 4/12 , 5/12 , 6/12 ) \gamma / k_b$ for $N=(1,2,3)$ pairs of molasses beams. In our case $N = 2$ and the result is in agreement with the simulation.

%figura_(8)
%
\section{Laser lens: Focusing}
\begin{figure}[H]
\begin{center}
\includegraphics[width=0.4\textwidth]{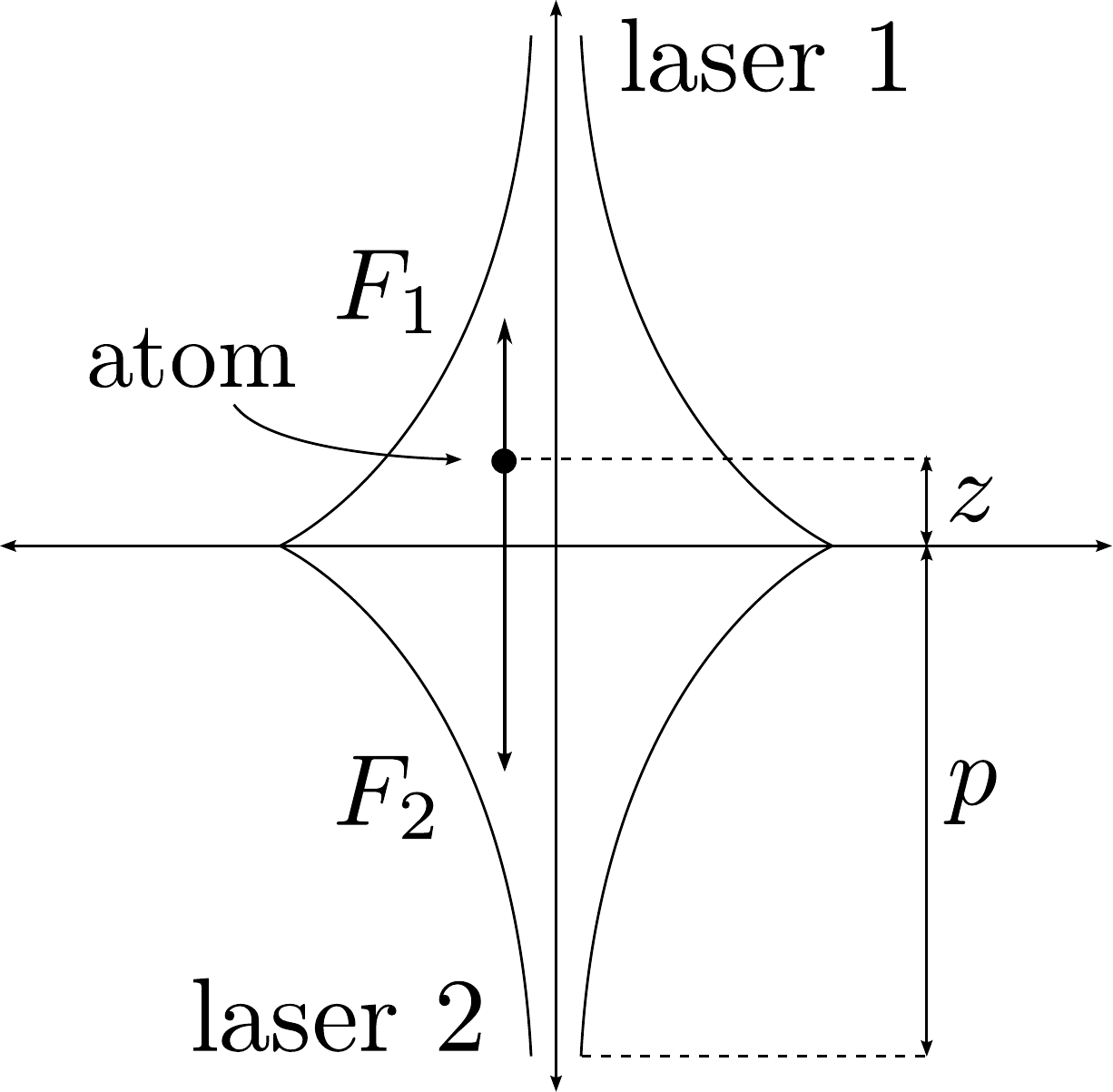}
\caption[Laser lens schematic]{\label{laserlens} Simplified schematic of a 1-D laser lens.}
\end{center}
\end{figure}

The 2-D OM induces a deflection in the atomic beam (see Fig.~\ref{zvsx_mol}), and collimates the beam down to the Doppler temperature in the transverse directions, but does not focus the atoms. This requires a position dependent force. Eq.~\ref{gauss} shows the possibility of introducing spatial dependence through the factor $w(z_i)$. By focusing the laser beams tightly at distance $p$ away from the central axis as shown in Fig.~\ref{laserlens}, we can create an intensity gradient over the interaction region and induce atomic beam focusing~\cite{balykin88}.

It is possible to achieve focusing by using a 2-D MOT as a magneto-optical lens~\cite{berthoud98,labeyrie99}. This gives a position dependent force from the magnetic field gradient that shifts the atomic resonances, allowing spatial compression. However, its setup carries undesirable consequences. The proximity of the 2-D MOT coils to the optical cavity, make accurate control of the magnetic fields in the cavity very difficult.

%figura_(9)

\begin{figure}[H]
\begin{center}
\includegraphics[width=0.8\textwidth]{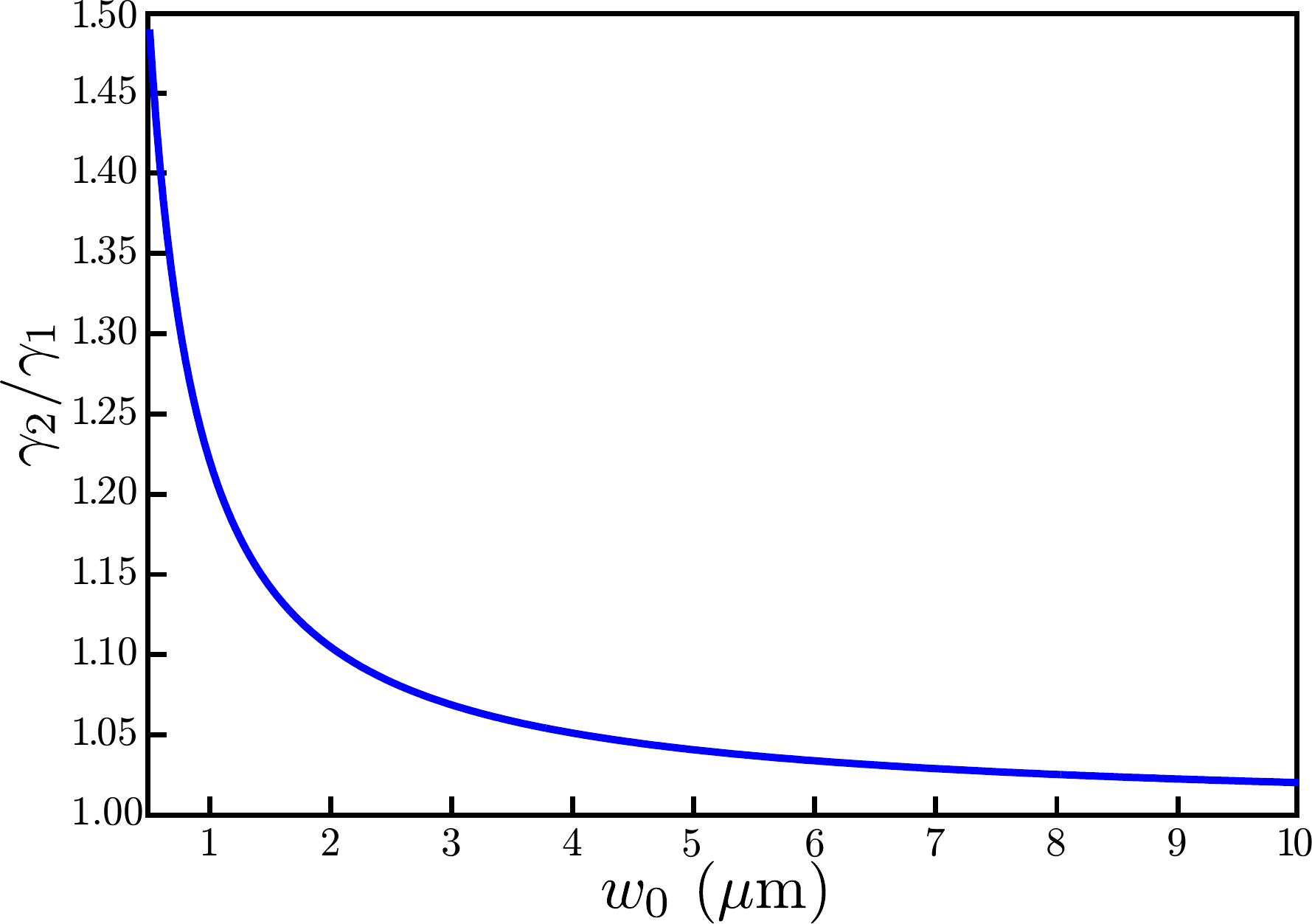}
\caption[Scattering rate imbalance]{\label{intengrad} Scattering rate ratio for an atom at $z = 1$ mm ($\vec{v} = 0$) in a laser lens with $s_0(0,0) = 3$ and $w(0) = 5$ mm.}
\end{center}
\end{figure}

Figure~\ref{laserlens} shows that when $\vec{v} = 0$, $F_1 - F_2 = - \hbar k \left( \gamma_{p,2} - \gamma_{p,1} \right) \neq 0$ away from the axis. We estimate how tightly focused the beams have to be to achieve a usable intensity imbalance. Fig.~\ref{intengrad} shows our results under the following assumptions. We take a fixed point for the calculation, at $z = 1$ mm. We choose this value because it represents a typical position of an atom in the beam as it exits the 2-D OM. The beam spot size at the center of the laser lens region is 5 mm and an intensity per beam at the center of $s_0 = 2$. We set $\vec{v} = 0$, to characterize the position dependent part of the interaction.

%figura_(10)

\begin{figure}[H]
\begin{center}
\includegraphics[width=0.45\textwidth]{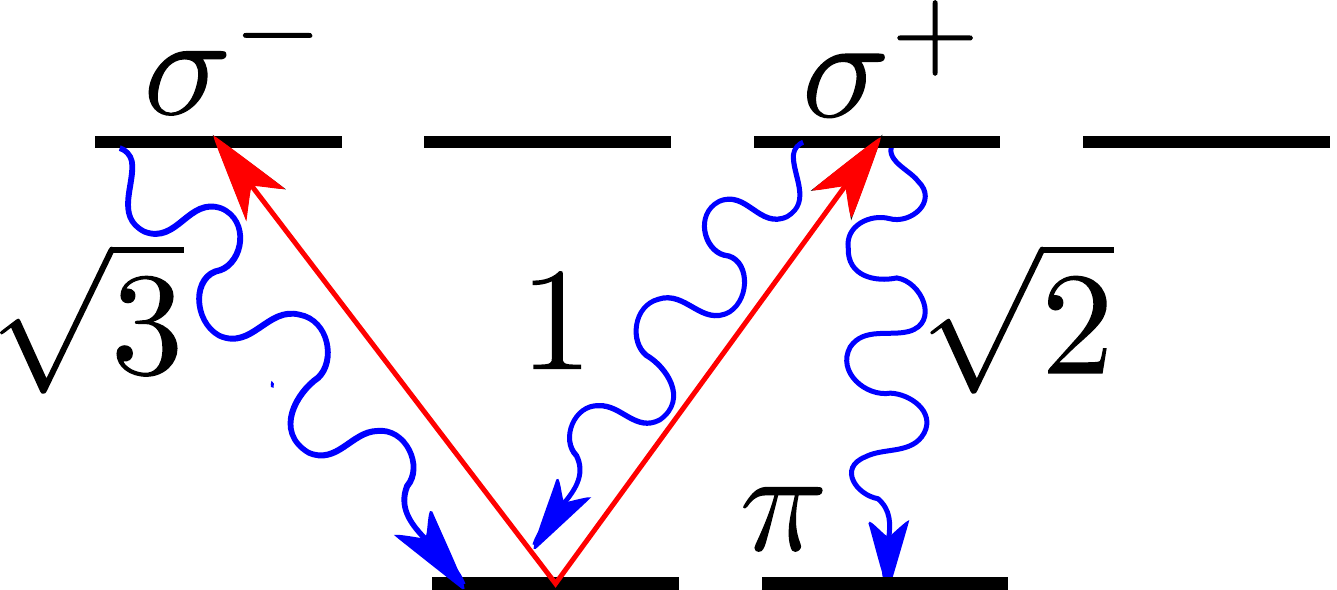}
\caption[6-level atom]{\label{6levels} (color online) 6-level atomic structure. Driven by $\sigma^{+}$ and $\sigma^{-}$ polarized light. Decays and excitations take place according the Clebsch-Gordan coefficients shown~\cite{metcalf99}.}
\end{center}
\end{figure}

These results indicate that a successful laser lens requires very tight focusing ($\sim$0.5 $\mu$m). We separate the deflection and the focusing, and elaborate a more sophisticated model for a two-dimensional laser that includes more of the atomic structure through a six level atom model correction~\cite{bouyer94} (See Fig.~\ref{6levels}).

%figura_(11)

The algorithm for the laser lens builds upon the one for the 2-D OM. The 6-level atom correction weighs the scattering rates per beam and spontaneous emission using the Clebsch-Gordan coefficients. We use two different quantization axes, $x$ and $y$, depending from which beam the atom most recently absorbed. This approach makes the code simpler.

Results from the simulation are complicated. They show focusing for very small waists. The refinements in the model (\textit{i.e.} 6-level correction) do not seem to loosen such tight requirements. We will continue to explore this, and evaluate its practical benefits.
\section{Conclusions}
We have demonstrated enhancements to increase the quality of our cavity and allow us to couple more atoms to its modes.

In cavity construction, we expect that careful experimental measurement of the transmissions of our mirrors have allowed us to make significantly more accurate predictions of finesse. Additionally, our progress in reducing mechanical vibrations via impedance mismatching of damping materials will help to ensure stable operation of the cavity.

We show encouraging simulation results for the implementation of 2-D optical molasses for the deflection and collimation of a Low Velocity Intense Source of atoms. Parameter explorations pointed at an optimal molasses laser beam detuning at $\delta / \gamma = -0.5$. By placing the cavity at a moderately short distance from the molasses, we showed that we can get up to 11 \% of atoms into the new cavity mode, in contrast with our simulation predictions for our current system of about 6 \%.

Despite initial evidence that the laser lens technique can achieve atomic focusing, we consider it highly difficult to implement in our experiment, due primarily to the very small laser beam waists required for its effectiveness ($<1$ $\mu$m).
\section*{Acknowledgments}
This work was supported by the National Science Foundation (NSF). We thank J. R. Ramos for his participation in early stages of the optical cavity building process. We are grateful to S. L. Rolston, H. J. Carmichael and M. Scholten for their stimulating discussions and guidance. Thanks to the \textsc{Root} developers and users for their invaluable advice.
\providecommand{\newblock}{}


\begin{thebibliography}{10}
\expandafter\ifx\csname url\endcsname\relax
  \def\url#1{{\tt #1}}\fi
\expandafter\ifx\csname urlprefix\endcsname\relax\def\urlprefix{URL }\fi
\providecommand{\eprint}[2][]{\url{#2}}

\bibitem{berman94}
Berman P~R (ed) 1994 {\em Cavity Quantum Electrodynamics\/} Advances in Atomic,
  Molecular, and Optical Physics (Boston: Academic Press) supplement 2

\bibitem{turchette95b}
Turchette Q~A, Hood C~J, Lange W, Mabuchi H and Kimble H~J 1995 {\em Phys. Rev.
  Lett.\/} {\bf 75} 4710--4713

\bibitem{cirac97}
Cirac J~I, Zoller P, Kimble H~J and Mabuchi H 1997 {\em Phys. Rev. Lett.\/}
  {\bf 78} 3221

\bibitem{gheri98}
Gheri K~M, Saavedra C, T{\"o}rm{\"a} P, Cirac J~I and Zoller P 1998 {\em Phys.
  Rev. A\/} {\bf 58} R2627--R2630

\bibitem{wilk07b}
Wilk T, Webster S~C, Kuhn A and Rempe G 2007 {\em Science\/} {\bf 317} 488

\bibitem{norris10}
Norris D~G, Orozco L~A, Barberis-Blostein P and Carmichael H~J 2010 {\em Phys.
  Rev. Lett.\/} {\bf {105}} {123602}

\bibitem{norris09a}
Norris D~G, Cahoon E~J and Orozco L~A 2009 {\em Phys. Rev. A\/} {\bf 80} 043830

\bibitem{terraciano09}
Terraciano M~L, {Olson~Knell} R, Norris D~G, Jing J, Fern{\'a}ndez A and Orozco
  L~A 2009 {\em Nat. Phys.\/} {\bf 5} 480--484

\bibitem{bishop08}
Bishop L~S, Chow J~M, Koch J, Houck A~A, Devoret M~H, Thuneberg E, Girvin S~M
  and Schoelkopf R~J 2008 {\em Nat. Phys.\/} {\bf 5} 105 -- 109

\bibitem{guerlin07}
Guerlin C, Bernu J, Del\'eglise S, Sayrin C, Gleyzes S, Kuhr S, Brune M,
  Raimond J~M and Haroche S 2007 {\em Nature\/} {\bf 448} 889--893

\bibitem{harochebook}
Haroche S and Raimond J~M 2006 {\em Exploring the Quantum: Atoms, Cavities, and
  Photons\/} 1st ed (Oxford University Press)

\bibitem{hennrich05}
Hennrich M, Kuhn A and Rempe G 2005 {\em Phys. Rev. Lett.\/} {\bf 94} 053604

\bibitem{lu96}
Lu Z~T, Corwin K~L, Renn M~J, Anderson M~H, Cornell E~A and Wieman C~E 1996
  {\em Phys. Rev. Lett.\/} {\bf 77} 3331

\bibitem{nasa_outgas}
NASA Outgassing data for selecting spacecraft materials
  \urlprefix\url{http://outgassing.nasa.gov/}

\bibitem{sorbothane}
Sorbothane \urlprefix\url{http://www.sorbothane.com/}

\bibitem{wang10}
Wang H and Iyanu G 2010 Mot-based continuous cold cs-beam atomic clock {\em
  Frequency Control Symposium (FCS), 2010 IEEE International\/} pp 454--458

\bibitem{nellessen89}
Nellessen J, M\"{u}ller J~H, Sengstock K and Ertmer W 1989 {\em J. Opt. Soc.
  Am. B\/} {\bf 6} 2149--2154

\bibitem{witte92}
Witte A, Kisters T, Riehle F and Helmcke J 1992 {\em J. Opt. Soc. Am. B\/} {\bf
  9} 1030--1037

\bibitem{ashkin70}
Ashkin A 1970 {\em Phys. Rev. Lett.\/} {\bf 25} 1321--1324

\bibitem{andres_sim_beam}
Cimmarusti A~D Simulation atomic beam control
  \urlprefix\url{http://terpconnect.umd.edu/~candres/projects.html}

\bibitem{metcalf99}
Metcalf H~J and Straten P 1999 {\em Laser Cooling and Trapping\/} (New York:
  Springer)

\bibitem{balykin88}
Balykin V~I, Letokhov V~S, Ovchinnikov Y~B and Sidorov A~I 1988 {\em Journal of
  Modern Optics\/} {\bf 35} 1734

\bibitem{berthoud98}
Berthoud P, Joyet A, Dudle G, Sagna N and Thomann P 1998 {\em Europhys.
  Lett.\/} {\bf 41} 141

\bibitem{labeyrie99}
Labeyrie G, Browaeys A, Rooijakkers W, Voelker D, Grosperrin J, Wanner B,
  Westbrook C and Aspect A 1999 {\em Eur. Phys. J. D\/} {\bf 7}(3) 341--350

\bibitem{bouyer94}
Bouyer P, Lemonde P, Dahan M~B, Michaud A, Salomon C and Dalibard J 1994 {\em
  Europhys. Lett.\/} {\bf 27} 569

\end{thebibliography}
\end{document}